astro-ph/9501109  31 Jan 1995

# Large Angle CMB Anisotropy in an Open Universe in the One-Bubble Inflationary Scenario

Kazuhiro Yamamoto[1], Misao Sasaki[2] and Takahiro Tanaka[1]

[1] Department of Physics, Kyoto University, Kyoto 606-01, Japan
[2] Department of Earth and Space Science, Osaka University, Toyonaka 560, Japan

## ABSTRACT

We consider an alternative scenario of inflation which can account for a spatially open universe. It is similar to the old inflation in which the bubble nucleation occurs in the sea of false vacuum, but differs from it in that the second slow rollover inflation occurs inside a nucleated bubble. Hence our observable universe is entirely contained in one nucleated bubble. The significance of the scenario is that apart from a variance due to model parameters it gives us a definite prediction on the spectrum of the primordial density perturbations, hence is observationally testable. Here we investigate the spectrum of CMB anisotropies on large angular scales. We find that the contribution from peculiar modes which never appear in the usual harmonic analysis is significant in the case $\Omega_0 \lesssim 0.1$.
*Subject headings:* cosmic microwave background — cosmology : theory

## I. INTRODUCTION

Inflationary cosmology has now become an integrated part of the standard model of the universe. However, predictions of inflationary cosmology do not always agree with observations. For example, some dynamical measurements indicate a low density universe $\Omega_0 < 1$ (Peebles 1993; Rattra & Peebles 1994a). Moreover, very recently, the distance to the Virgo cluster of galaxies was determined by using Cepheids as $17.1 \pm 1.8$ Mpc, which implies the Hubble parameter of $H_0 = 80 \pm 17 \text{km/s/Mpc}$ (Freedman et al. 1994). Requiring this value to be consistent with the age of globular clusters, we are then forced to conclude that our universe is either spatially open or cosmological constant-dominated. A natural prediction of the standard inflationary scenario is that the universe is spatially flat. Hence if we respect these observations we have to assume a cosmological constant-dominated universe. However, in this paper, we take a different view and consider a possible scenario of an open universe in the context of inflationary cosmology.

Naively one might think that it is possible to realize an open universe by suitably adjusting the period of inflation. However, as discussed by Kashlinski, Tkachev and Frieman (Kashlinski et al. 1994), it is very difficult to realize a homogeneous and isotropic universe within the standard inflationary universe scenario. This is because the horizon and flatness problems, which are two different problems in nature, happen to be closely related in the standard scenario, *i.e.*, they are explained simultaneously by the large *e*-folding number of the inflationary expansion.

It is clear that this difficulty can only be resolved by separating the horizon and flatness problems and solving each with a different mechanism. One-bubble inflationary model is one of such scenarios. Originally this idea was proposed by Gott (Gott 1982; Gott & Statler 1983). In this scenario, a bubble is nucleated due to quantum tunneling in the sea of false vacuum. The bubble nucleation process is described by an $O(4)$-symmetric bounce solution (Coleman 1977; Coleman & De Luccia 1980), which is a solution of the field equation in the Euclidean spacetime. The subsequent motion of a bubble after nucleation is given by the analytic continuation of the bounce solution to the Lorentzian spacetime. Hence the spacetime has the $O(3,1)$-invariance owing to the $O(4)$-symmetry of the bounce solution. In other words, the interior of a bubble is a homogeneous and isotropic open universe, which solves the horizon problem. Then if another inflation occurs subsequently in the bubble for a suitable amount of time, the flatness with $\Omega_0 \sim 0.1$ of the universe can be explained without conflicting with the homogeneity and isotropy of the universe.

This scenario can be realized by introducing two or more different mass scales into the scalar field potential (Guth & Weinberg 1983). As a typical example, we consider a scalar field with a new inflation type potential with a high potential barrier (see Figure 1). For such a potential, the first inflation in the false vacuum provides a coherently homogeneous scalar field due to the de Sitter expansion. Then, a high potential barrier guarantees an exponentially small bubble nucleation rate so that bubble collisions can be neglected. Thus the bubble nucleation process can be regarded as the creation of a homogeneous and isotropic universe with negative spatial curvature. Then the second inflation of slow rollover type in the bubble explains $\Omega_0 \sim 0.1$. The subsequent evolution of universe after the second inflation is the same as the usual inflation scenario.

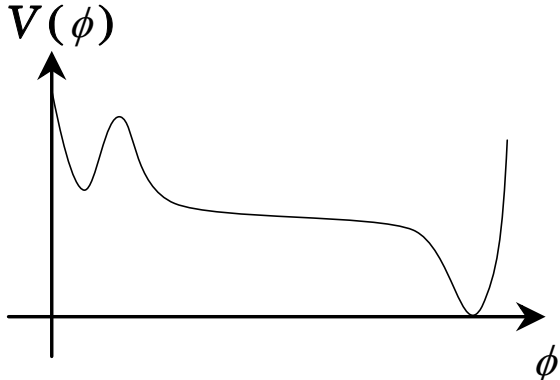

**Fig. 1** A schematic picture of the potential which realizes the one-bubble inflationary scenario.

Within the context of this new scenario, the important issue to be addressed is the origin of the density perturbations for the cosmic structure formation. As in the standard inflationary scenario, we expect the quantum fluctuations of the scalar field to give rise to the cosmic density perturbations. There have been many papers devoted to the investigation of the perturbations in an open universe. However most of them are always based on some *ad hoc* assumption for the initial perturbation spectrum due to the lack of a definite scenario. In contrast, our scenario is definite enough that one is able to predict the initial perturbation spectrum as quantitative as one desires in principle.

In our open inflationary scenario, the quantum fluctuations of the field originate from two different parts. One is the fluctuations caused during the bubble nucleation process (Rubakov 1984; Vachaspati & Vilenkin 1991; Tanaka et al. 1994; Sasaki et al. 1993a; Tanaka & Sasaki 1994; Yamamoto et al. 1995; Hamazaki et al. 1995). The other is the usual vacuum fluctuations which are imprinted during the first and second inflationary phases. In this paper, as a first attempt to quantify the predictions of the scenario, we investigate the consequent CMB anisotropies on large angular scales by assuming that the fluctuations caused during the bubble nucleation are negligible.

This paper is organized as follows. In section 2, we first briefly review the difficulty to construct an open universe scenario in the standard inflation, then describe the one-bubble inflationary model and examine the condition for this model to be successful. In section 3, we consider the quantum fluctuations of the scalar field and give the spectrum of curvature perturbation based on our recent results on quantum field theory in the open de Sitter spacetime (Sasaki et al. 1994). Then the temperature fluctuations on large angular scales are calculated. Finally section 4 is devoted to discussions. We take the units $c = 1$ and $\hbar = 1$.

## II. ONE-BUBBLE INFLATIONARY UNIVERSE

The most attractive feature of the inflationary universe scenario is that it can elegantly explain the homogeneity, isotropy and flatness of the universe. However, it is possible that our universe today may not be quite flat but open with $\Omega_0 \sim 0.1$. If so, it would be a fatal blow against the standard inflationary universe scenario as discussed by Kashlinsky, Tkachev and Frieman (Kashlinski et al. 1994). The reason is that the homogeneity and isotropy of the universe and the flatness are intertwined in the standard scenario of inflation; they are simultaneously realized by a large amount of expansion of the universe. To make the motivation for one-bubble inflationary scenario more manifest, let us summarize this point.

The Friedmann equation for an open universe may be written as

$$1 = \left(1 - \Omega(t)\right) H^2(t) a^2(t), \tag{2.1}$$

where $a(t)$ is the scale factor normalized to be the spatial curvature radius, $H(t) := \dot{a}(t)/a(t)$ is the Hubble parameter, and $\Omega(t)$ is the density parameter. In the standard scenario of inflation, a comoving region which coincides with the whole observable universe today (*i.e.*, the present Hubble horizon size) was equal to the Hubble horizon size during the inflationary stage. Let us denote the proper size of this comoving region by $L(t)$ and the epoch at which $L(t) = 1/H(t)$ during the inflation by the subscript '$h$'. We then have

$$H_0 = \frac{1}{L_0} = \frac{a_h}{a_0}\frac{1}{L_h} = \frac{a_h}{a_0} H_h, \tag{2.2}$$

which implies

$$H_0 a_0 = H_h a_h, \tag{2.3}$$

where the subscript '0' denotes the present time. On the other hand, from the Friedmann equation (2.1), we have

$$1 = (1 - \Omega_0)(H_0 a_0)^2 = (1 - \Omega_h)(H_h a_h)^2. \tag{2.4}$$

We immediately see that Eqs. (2.3) and (2.4) are consistent only if $\Omega_0 = \Omega_h$. Now, according to the standard inflation scenario, the homogeneity and isotropy of the universe are realized only for a comoving region whose size exceeded the horizon size well after the inflation begun. This implies $\Omega_h = 1$ to a high accuracy, hence $\Omega_0 = 1$. Putting it in another way, if the epoch $t = t_h$ were to be at the beginning of inflation, one could have $\Omega_h$ appreciably smaller than unity, but one could not expect the region of size $L_h = 1/H_h$ to be sufficiently homogeneous and isotropic, since such an anticipation would be against the spirit of inflation. By analyzing the contribution to the CMB anisotropy from the superhorizon scale perturbations, it has been shown that only the value $|1 - \Omega_0| < 10^{-5}$ is allowed to match with the observed smoothness of CMB as long as we follow the standard inflation scenario (Kashlinski et al. 1994).

Thus, to realize a sufficiently homogeneous and isotropic open universe, we need a different scenario other than the standard inflation. As mentioned in Sec.1, the one-bubble inflationary universe model is one of such possibilities. This model assumes a scalar field which has a potential of new inflation type but with a high potential barrier (Fig. 1). Initially the scalar field is trapped in the false vacuum minimum. If the potential barrier is high enough, the decay rate is exponentially suppressed and the universe expands exponentially for a sufficiently long period before the bubble nucleation occurs. Then when a bubble nucleates it occurs in the smooth background of de Sitter spacetime. In this case, the bubble nucleation is mediated by the $O(4)$-symmetric bounce solution of the Euclidean field equation. Consequently the nucleated bubble in the real Lorentzian region has the $O(3,1)$-symmetry and the interior of the bubble is an open universe having this symmetry (Gott 1982; Sasaki et al. 1993b), that is, it is a homogeneous and isotropic universe with negative spatial curvature.* Then the scalar field starts rolling down the hill of the new inflation type potential as in the standard inflationary scenario. In this manner the homogeneity and isotropy of the universe are guaranteed at the onset of the second inflation contrary to the standard scenario. This allows us to satisfy the condition $\Omega_h = \Omega_0 \sim 0.1$ without violating the homogeneity and isotropy of the comoving region of size $L(t)$ by suitably adjusting the model parameters. It should be noted that this adjustment of the model parameters is a kind of fine-tuning but it is *not* a fine-tuning of a kind that is necessary in the standard inflationary scenario; in the latter case one is required to fine-tune the homogeneity and isotropy of the universe which cannot be controlled by the model parameters while it is a controllable fine-tuning in our case.

Then the crucial condition for this one-bubble scenario to be successful is that the probability that a nucleated bubble collides with another nucleated bubble is small enough so that the universe inside a bubble can survive until today. Interestingly, the fulfillment of this very condition was the reason why the old inflationary universe scenario was abandoned, namely the most part of the spacetime continues to inflate forever (Guth & Weinberg 1981). In other words, as long as we have a non-inflationary stage before the first false vacuum inflation, the assumed sufficiently small nucleation rate automatically guarantees the smallness of the collision probability. Of course the precise value of the probability depends on the precise scenario of the pre-inflationary stage. Quantitative evaluation of the collision probability for a model in which the inflationary stage of the universe is approximated by the spatially flat chart of de Sitter space has been done by Gott & Statler (1983). However the result will be qualitatively the same for any other model; it is essentially the bubble nucleation rate per unit Hubble time per unit Hubble volume. One might worry what would happen if there were no pre-inflationary stage, that is, if the inflation were eternal to the past. However, Borde and Vilenkin have shown that the past-eternal inflation is impossible (Borde & Vilenkin 1994). Hence inflation must have a beginning and once it has a beginning the small collision probability is guaranteed by the small nucleation rate.

Now let us consider the condition that $\Omega_0$ is neither very close to unity nor too much smaller than it. It has been discussed by Bucher, Goldhaber, & Turok (1994). We essentially repeat their argument. The dynamics of the scalar field inside a bubble is described by

$$\ddot{\phi} + 3\frac{\dot{a}}{a}\dot{\phi} + V'(\phi) = 0, \tag{2.5}$$

$$\dot{a}^2 = 1 + \frac{8\pi a^2}{3m_{pl}^2}\left(\frac{1}{2}\dot{\phi}^2 + V\right), \tag{2.6}$$

where $m_{pl}$ is the Planck mass. The initial conditions at $t = 0$ are given by the analytic continuation of the bounce solution (Coleman & De Luccia 1980),

$$\begin{aligned}\dot{a} &= 1, &\dot{\phi} &= 0,\\ a &= 0, &\phi &= \phi_b,\end{aligned} \tag{2.7}$$

where $\phi_b$ is the value of $\phi$ at the center of the $O(4)$-symmetric bounce. We assume the potential $V$ is sufficiently flat around $\phi = \phi_b$ so that the scalar field satisfies the potential-dominated slow-rolling condition, *i.e.*, $\dot{\phi}^2/2 \ll V$ and $|\ddot{\phi}| \ll |3H\dot{\phi}|$, which reduce to the conditions for the potential as (Kolb & Turner 1990)

$$\frac{m_{pl}^2 V''}{24\pi V} \ll 1, \quad \frac{m_{pl}^2 V'^2}{48\pi V^2} \ll 1. \tag{2.8}$$

In the slow-rolling limit $\dot{\phi} \to 0$, the solution is given by

$$a(t) \approx \frac{1}{H_*}\sinh H_* t, \quad \phi(t) \approx \phi_b; \quad H_*^2 := \frac{8\pi V(\phi_b)}{3m_{pl}^2}. \tag{2.9}$$

Using the slow-rolling equation of motion for $\phi$,

$$\dot{\phi} = -\frac{V'}{3H}, \tag{2.10}$$

---

*Here we are focusing on the lowest order (classical) picture of inflation and neglecting the quantum fluctuation of the scalar field, which will be the topic of the next section.

we find the approximate solution (2.9) is valid during the era $H_*t \ll 8\pi V^2/m_{pl}^2 V'^2$. Then the second of the slow-rolling conditions (2.8) implies it is valid at least for the stage $H_*t$ is not too much larger than unity. In other words, the approximate solution (2.9) is valid for the entire stage during which the curvature effect may be non-negligible. On the other hand for $H_*t \gtrsim 1$, we may neglect the curvature and the solution is given by that in the case of the standard inflationary scenario. For simplicity, we divide these two eras at the time when $H_*a = 1$ and denote it by $t_*$, i.e., $\sinh H_*t_* = 1$. We note that, since we assume $\Omega_0$ to be not very close to unity, the time $t = t_h$ at which the comoving scale $L(t)$ of the present horizon size came out of the horizon should be in the range $t_h < t_*$.

Now the total number of $e$-folds $N_h$ from $t = t_h$ to the end of inflation, whose epoch we denote by $t = t_e$, is evaluated as

$$N_h = \int_{t_h}^{t_e} H\, dt = \int_{t_h}^{t_*} H\, dt + \int_{t_*}^{t_e} H\, dt$$
$$\approx \frac{1}{2} \ln(\Omega_0^{-1} - 1) + \frac{8\pi}{m_{pl}^2} \int_{\phi_e}^{\phi_*} \frac{V}{V'} d\phi, \quad (2.11)$$

where $\phi_e = \phi(t_e)$, $\phi_* = \phi(t_*) \approx \phi_b$ and we have neglected the curvature term at $t \geq t_*$. According to our assumption of a sufficient flat potential, we may approximate the value $\phi_*$ by $\phi_b$. We require $N_h$ to be just a value such that the comoving size $L(t)$ coincides with the horizon at present. Hence $N_h$ is the same as the minimum number of $e$-folds to solve the horizon problem in the standard scenario (Kolb & Turner 1990):

$$N_h = 53 + \frac{2}{3} \ln\left(\frac{V_e^{1/4}}{10^{14} \text{GeV}}\right) + \frac{1}{3} \ln\left(\frac{T_{RH}}{10^{10} \text{GeV}}\right), \quad (2.12)$$

where $V_e$ is the value of the potential at $t = t_e$ and $T_{RH}$ is the reheating temperature. From Eq. (2.11) with $\phi_* = \phi_b$, we now have the condition on the potential as

$$\int_{\phi_e}^{\phi_b} \frac{V}{V'} \frac{d\phi}{m_{pl}^2} = \frac{1}{16\pi} \ln\left(\frac{e^{2N_h}}{\Omega_0^{-1} - 1}\right), \quad (2.13)$$

where $N_h$ is the value given by Eq. (2.12).

As noted before, although this is a fine-tuning of the parameters of the potential as well as of the strength of the coupling to other matter fields, it is important that this is a controllable fine-tuning. Further, as argued by Bucher, Goldhaber, & Turok (1994), the required value of $\phi_b$ will be of Planck mass order for polynomial potentials, which is not unreasonable.

## III. CURVATURE PERTURBATION AND CMB ANISOTROPY

In this section, we evaluate the curvature perturbation spectrum induced by the quantum fluctuations of the scalar field in the one-bubble inflationary scenario and calculate the resultant CMB anisotropies on large angular scales. Recently some authors have investigated the same problem in open inflationary universe models (Bucher et al. 1994; Lyth & Stewart 1990; Rattra & Peebles 1994b; Kaminkowski et al. 1994). However, except for the work by Bucher, Goldhaber, & Turok (1994), they had to give the initial spectrum of quantum fluctuations of the inflaton field by hand. In contrast, in the present one-bubble scenario, we have a definite prediction on the spectrum of quantum fluctuations, which may be calculable in principle to any desired accuracy, given the sufficient information of the model parameters. For example, under some additional assumptions for simplification of the model, Bucher, Goldhaber, & Turok (1994) have derived the spectrum of curvature perturbations in the one-bubble open inflationary scenario. Here we consider a simple situation, which is different from what Bucher, Goldhaber and Turok have assumed, but which retains an essential feature of the one-bubble scenario.

Because the bubble nucleation is assumed to occur sufficiently rarely, the quantum state of any scalar field at the false vacuum may be well approximated by the de Sitter-invariant Euclidean vacuum. Then when a bubble nucleates, the quantum state inside the bubble will be also the de Sitter-invariant Euclidean vacuum, provided that the effect of the tunneling process on the quantum state is small. In particular, this is true for fields which do not interact with the tunneling field. This assumption will fail for the scalar field which describes the fluctuation of the tunneling field itself, which will be the case if the inflaton field for the second inflation inside the bubble is the same as the tunneling field. However, since this effect from the tunneling process seems to depend very much on the precise form of the interaction between (the fluctuation of) the inflaton field and the tunneling (Yamamoto et al. 1994; Hamazaki et al. 1995), we defer this issue to future work and assume here that the quantum state inside the bubble is also well approximated by the de Sitter-invariant Euclidean vacuum.

The quantum field theory in the de Sitter spacetime has been well investigated in the flat and closed charts (Birrell & Davies 1982; Bunch & Davies 1978; Chernikov & Tagirov 1968). To find the spectrum of fluctuations in the open universe, however, we need to describe the Euclidean vacuum in terms of the mode functions associated with the open chart of the de Sitter space, i.e., where the metric is written in the form

$$ds^2 = -dt^2 + a^2(t)\left(d\chi^2 + \sinh^2\chi\, d\Omega_{(2)}^2\right), \quad (3.1)$$

where $a(t)$ is the scale factor $a(t) = H^{-1} \sinh Ht$. Here and in what follows, $H$ is not the Hubble parameter but is a constant representing the curvature scale of de Sitter space. Recently we have succeeded in solving this problem (Sasaki et al. 1994). According to the result, a quantized minimally coupled scalar field $\hat{\phi}(x)$ with mass $m$ ($m^2 < 2H^2$) in the de Sitter-invariant Euclidean vacuum is described in the second quantization manner, as

$$\hat{\phi}(t,\chi,\Omega) = \int_0^\infty dp \sum_{\sigma,l,m} w_{p,\sigma}(t) Y_{plm}(\chi,\Omega) \hat{a}_{p\sigma lm}$$
$$+ \sum_{l,m} v_{(*)lm}(t,\chi,\Omega) \hat{a}_{(*)lm} + \text{h.c.} , \quad (3.2)$$

where

$$w_{p,\sigma}(t) = \sqrt{\frac{\pi |\Gamma(\nu'+1+ip)|^2}{4(\cosh \pi p - \sigma \cos \pi \nu')}}$$
$$\times \frac{1}{2\sinh \pi p\, a(t)} \left[ \frac{e^{\pi p} - \sigma e^{-i\pi \nu'}}{\Gamma(\nu'+1+ip)} P_{\nu'}^{ip}(\cosh Ht) \right.$$
$$\left. - \frac{e^{-\pi p} - \sigma e^{-i\pi \nu'}}{\Gamma(\nu'+1-ip)} P_{\nu'}^{-ip}(\cosh Ht) \right],$$
$$(3.3)$$

$$Y_{plm}(\chi,\Omega) = \left|\frac{p\Gamma(ip+l+1)}{\Gamma(ip+1)}\right| \frac{P_{ip-1/2}^{-l-1/2}(\cosh \chi)}{\sqrt{\sinh \chi}} Y_{lm}(\Omega)$$
$$=: W_{pl}(\chi) Y_{lm}(\Omega),$$
$$(3.4)$$

$$v_{(*)lm}(t,\chi,\Omega) = \sqrt{\nu'^2 \Gamma(2\nu') \Gamma(\nu'+l+1)\Gamma(-\nu'+l+1)}$$
$$\times \frac{P_{\nu'}^{-\nu'}(\cosh Ht)}{a(t)} \frac{P_{\nu'-1/2}^{-l-1/2}(\cosh \chi)}{\sqrt{\sinh \chi}} Y_{lm}(\Omega),$$
$$(3.5)$$

and $\sigma$ takes $\pm 1$, $\Gamma(z)$ is the Gamma function, $P_\mu^\nu(z)$ is the Legendre function of the first kind, $Y_{plm}(\chi,\Omega)$ is the orthonormal harmonics on a three dimensional unit hyperboloid, $\nu' := \sqrt{9/4 - m^2/H^2} - 1/2$. Then the Euclidean vacuum is annihilated by $\hat{a}_\Lambda$, where $\Lambda = \{p\sigma lm\}$ or $\{(*)lm\}$. A peculiar feature of the representation (3.2) is the existence of a set of discrete modes (3.5), which are modes with an imaginary value of $p$ ($p = i\nu'$) and hence are unnormalizable on the hyperboloid but are nonsingular. It has been clarified that the existence of these modes is essential to recover the de Sitter invariance for a scalar field with $m^2 < 2H^2$ (Sasaki et al. 1994). Very recently, necessity of the unnormalizable modes with $-1 < p^2 < 0$ for complete description of a random field in the open universe has been also discussed by Lyth & Woszczyna (1995).

In the small mass limit $m^2 \ll H^2$, Eq.(3.3) reduces to

$$w_{\sigma,p}(t) = \frac{-1}{\sqrt{8p(p^2+1)\sinh \pi p}}$$
$$\times \left[ e^{\pi p/2}(ip+\coth \eta)e^{-ip\eta} \right.$$

$$\left. + \sigma e^{-\pi p/2}(ip - \coth \eta) e^{ip\eta} \right] \frac{1}{a(\eta)}, \quad (3.6)$$

and Eq.(3.5) to

$$v_{(*)lm}(t,\chi,\Omega) = \frac{H}{2}\sqrt{\Gamma(l+2)\Gamma(l)} \frac{P_{1/2}^{-l-1/2}(\cosh \chi)}{\sqrt{\sinh \chi}} Y_{lm}(\Omega),$$
$$(l \geq 1)$$
$$=: \frac{H}{2} W_{(*)l}(\chi) Y_{lm}(\Omega),$$
$$(3.7)$$

where we have used the formula, $P_\nu^{-\nu}(z) = 2^{-\nu}(z^2-1)^{\nu/2}/\Gamma(1+\nu)$, and the conformal time is introduced by $\sinh Ht = -1/\sinh \eta$.

According to Lyth & Stewart (1990), the curvature perturbation of comoving hypersurfaces in an open inflationary universe is still given by $\mathcal{R}_\Lambda \simeq -(H/\dot{\phi})\delta \phi_\Lambda$ on superhorizon scales, where $\delta \phi_\Lambda$ is defined on flat hypersurfaces on which the effect of the metric perturbation is minimal, hence can be ignored. As long as $|1-\Omega| \ll 1$ which is true except for the very first moment of the bubble nucleation and the very recent stage after decoupling, $\mathcal{R}_\Lambda$ on scales greater than the horizon remains constant in time. For modes which come inside the horizon after the universe becomes matter-dominated, the gravitational potential perturbation $\Psi_\Lambda$ is related to the curvature perturbation $\mathcal{R}_\Lambda$ as $\Psi_\Lambda \simeq (3/5)\mathcal{R}_\Lambda$[†] (Bardeen 1980; Kodama & Sasaki 1984). Hence we may evaluate $\delta \phi_\Lambda$ in terms of the late time behavior of the mode functions given above at the inflationary stage and take the resulting $\Psi_\Lambda$ as the initial amplitude of the potential perturbation in the matter-dominated universe. Therefore, identifying thus obtained $\Psi_\Lambda$ with $\Psi_\Lambda(\eta=0)$, where now $\eta$ is the conformal time for the matter-dominated stage at which the scale factor is given by $a(\eta) = \cosh \eta - 1$, we obtain

$$\Psi_{p\sigma lm}(\eta=0,\chi,\Omega) = -\frac{3H^2}{5\dot{\phi}} \frac{(e^{\pi p/2} - \sigma e^{-\pi p/2})}{\sqrt{8p(p^2+1)\sinh \pi p}} Y_{plm}(\chi,\Omega)$$
$$=: \Psi_{p\sigma}(\eta=0) Y_{plm}(\chi,\Omega), \quad (3.8)$$

$$\Psi_{(*)lm}(\eta=0,\chi,\Omega) = -\frac{3H^2}{5\dot{\phi}} \frac{1}{2} W_{(*)l}(\chi) Y_{lm}(\Omega)$$
$$=: \Psi_{(*)}(\eta=0) W_{(*)l}(\chi) Y_{lm}(\Omega), \quad (3.9)$$

where $H^2/\dot{\phi}$ is to be evaluated at the inflationary stage at which $1-\Omega \ll 1$. At the matter-dominated stage, the potential perturbation evolves as (Weinberg 1972),

---

[†]The quantities which we have called $\mathcal{R}_\Lambda$ and $\Psi_\Lambda$ here are called $\phi_m$ and $\Phi_H$, respectively, in the cited reference (Bardeen 1980).

$$\Psi_\Lambda(\eta, \chi, \Omega) = \Psi_\Lambda(\eta = 0, \chi, \Omega) F(\eta), \qquad (3.10)$$

where

$$F(\eta) = 5 \frac{\sinh^2 \eta - 3\eta \sinh \eta + 4 \cosh \eta - 4}{(\cosh \eta - 1)^3}. \qquad (3.11)$$

The potential perturbations give rise to CMB anisotropies on large angular scales, the so-called Sach-Wolfe effect. In a matter-dominated open universe, they are given by (Gouda et al. 1991; Kaminkowski & Spergel 1994)

$$\Theta_{p\sigma lm}(\eta_0, \chi = 0, \Omega) = \Psi_{p\sigma}(\eta = 0) \left\{ \frac{1}{3} F(\eta_{LS}) W_{pl}(\eta_0 - \eta_{LS}) + 2 \int_{\eta_{LS}}^{\eta_0} \frac{dF(\eta')}{d\eta'} W_{pl}(\eta_0 - \eta') \right\} Y_{lm}(\Omega), \qquad (3.12)$$

$$\Theta_{(*)lm}(\eta_0, \chi = 0, \Omega) = \Psi_{(*)}(\eta = 0) \left\{ \frac{1}{3} F(\eta_{LS}) W_{(*)l}(\eta_0 - \eta_{LS}) + 2 \int_{\eta_{LS}}^{\eta_0} \frac{dF(\eta')}{d\eta'} W_{(*)l}(\eta_0 - \eta') \right\} Y_{lm}(\Omega), \qquad (3.13)$$

where $\Theta_{p\sigma lm}$ and $\Theta_{(*)lm}$ are the spherical harmonic components of the CMB anisotropies due to the continuous and discrete modes, respectively, and $\eta_0$ and $\eta_{LS}$ are the present conformal time and the conformal time of the last scattering, respectively.

As conventionally done, we write the temperature autocorrelation function in the form,

$$C(\alpha) = \frac{1}{4\pi} \sum_l (2l+1) C_l P_l(\cos \alpha). \qquad (3.14)$$

Then the multipole moment $C_l$ in the present case is given by

$$C_l = C_{(n.m.)l} + C_{(*)l}, \qquad (3.15)$$

where

$$C_{(n.m.)l} = \left(\frac{3H^2}{5\dot\phi}\right)^2 \int_0^\infty dp \frac{\coth \pi p}{2p(p^2+1)} \left\{ \frac{1}{3} F(\eta_{LS}) W_{pl}(\eta_0 - \eta_{LS}) + 2 \int_{\eta_{LS}}^{\eta_0} \frac{dF(\eta')}{d\eta'} W_{pl}(\eta_0 - \eta') \right\}^2, \qquad (3.16)$$

$$C_{(*)l} = \left(\frac{3H^2}{5\dot\phi}\right)^2 \frac{1}{4} \left\{ \frac{1}{3} F(\eta_{LS}) W_{(*)l}(\eta_0 - \eta_{LS}) + 2 \int_{\eta_{LS}}^{\eta_0} \frac{dF(\eta')}{d\eta'} W_{(*)l}(\eta_0 - \eta') \right\}^2. \qquad (3.17)$$

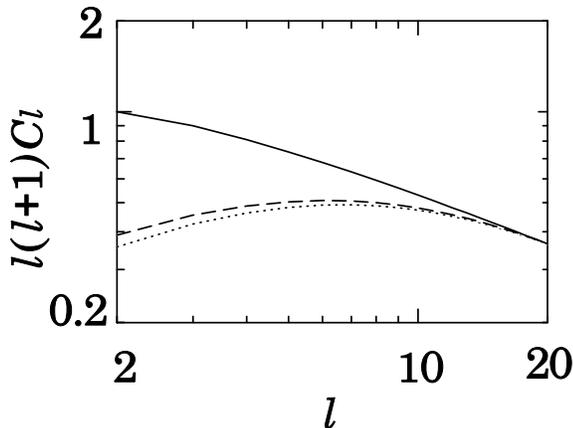

**Fig. 2** The multipoles of the temperature autocorrelation function, $C_l$, are plotted for $2 \leq l \leq 20$ by the bold line, setting $\Omega_0 = 0.05$. The dashed line is the contribution only from the continuous modes, $C_{(n.m.)l}$, and the dotted line shows the case in which the quantum state is in the conformal vacuum, $C_{(CV)l}$.

To find the multipole coefficients $C_l$, we have performed numerical integrations. The results are shown by the bold lines in Figures 2-6 for $2 \leq l \leq 20$. The amplitudes of $C_l$ are normalized by $C_2$. For reference, the dashed lines show the contributions only from the continuous modes, $C_{(n.m.)l}$. In order to compare with the previous analyses (Lyth & Stewart 1990; Rattra & Peebles 1994b; Kaminkowski et al. 1994), we have calculated the multipole coefficients when the scalar field is in a conformal vacuum state in open de Sitter space. The multipole coefficients in that case, which we write as $C_{(CV)l}$, are given in the same form of the right hand side of Eq.(3.16) but with $\coth \pi p$ removed. The results are shown by the dotted line in figures. The amplitudes are normalized by $C_2$.

Figure 2 and 3 show the cases $\Omega_0 = 0.05$ and $\Omega_0 = 0.1$, respectively. As expected, the discrete modes contribute to only small $l$ ($l \lesssim 10$). In these cases, the maximum contributions are about 60 and 30 percent, respectively, for $l = 2$. The differences between $C_{(n.m.)l}$ and $C_{(CV)l}$ are of order of a few percent at $l \lesssim 10$ for both cases. Figure 4 shows the case $\Omega_0 = 0.2$, for which the contribution from the discrete modes is less than the case $\Omega_0 = 0.1$. Figures 5 and 6 show the cases $\Omega_0 = 0.4$ and $\Omega_0 = 0.5$, respectively. In these cases, the contributions from the discrete modes are negligible. In addition, as $\Omega_0$ increases the spectrum becomes flatter, as naturally expected.

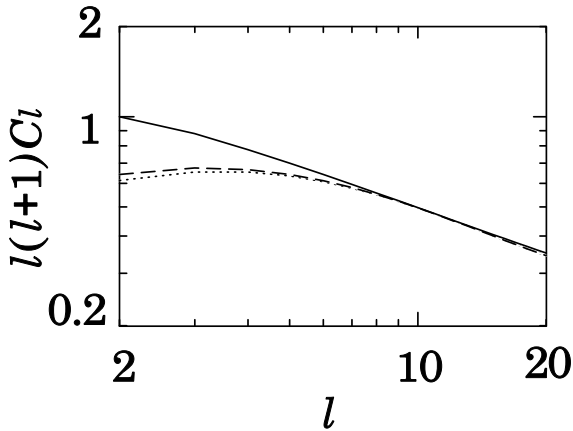

**Fig. 3** The same as figure 2 but for $\Omega_0 = 0.1$.

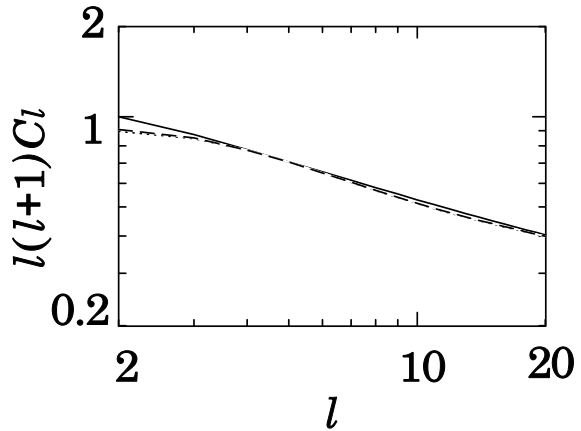

**Fig. 4** The same as figure 2 but for $\Omega_0 = 0.2$.

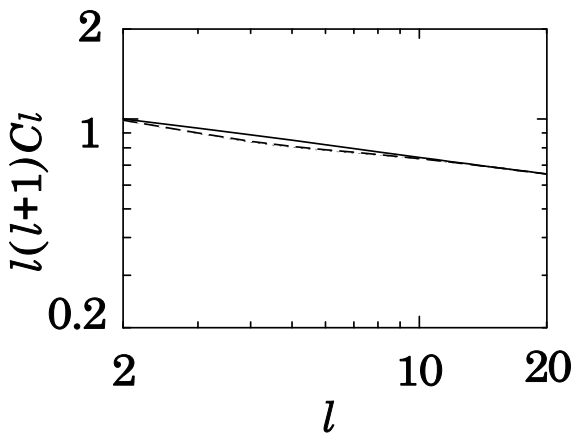

**Fig. 5** The same as figure 2 but for $\Omega_0 = 0.4$.

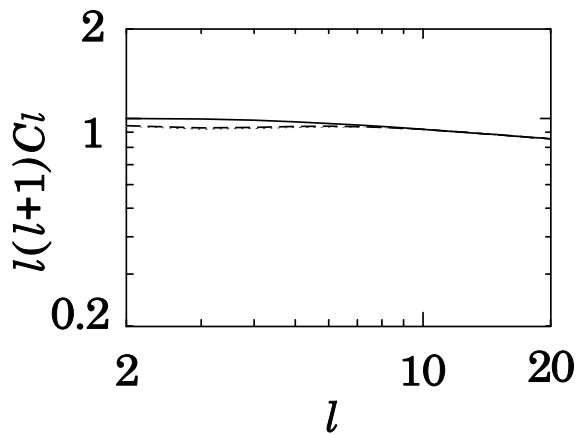

**Fig. 6** The same as figure 2 but for $\Omega_0 = 0.5$.

## IV. DISCUSSION

In this paper, we have considered the scenario which predicts an open universe $\Omega_0 \sim 0.1$. Such a scenario can be realized if we assume a scalar field with a new inflation type potential with a high potential barrier. Because of the first inflation in false vacuum, which provides a coherently homogeneous scalar field, and the small decay rate due to the high potential barrier, the nucleation of one bubble is regarded as creation of a homogeneous and isotropic open universe. Then the second inflation explains the large entropy of the universe or the almost flatness of the universe. Thus the mechanism to explain the homogeneity and isotropy of the universe and the entropy problem are separated, in contrast to the usual inflation scenario.

In this scenario, one can predict the spectrum of the initial perturbations for cosmic structure formation as quantitative as one desires in principle, following the usual picture that the initial density perturbations are generated by quantum fluctuations. Here, as a fist step of investigation, we have assumed that the quantum state of the scalar field inside the bubble is in the Euclidean vacuum state, and have not considered the effect of field excitation due to the bubble nucleation process. With this assumption, we have investigated the initial condition of the cosmological perturbation on superhorizon scale, focusing on the large angle CMB anisotropies.

By using the relation that connects the curvature perturbation with the fluctuation of the scalar field, which has been discussed by Lyth & Stewart (1990), we have evaluated the ensemble average of the multipole coefficients of the CMB anisotropies on large angular scales. We have obtained them by numerical integrations. The point to note is the contribution from the discrete modes which are not normalizable but are nonsingular on a hyperboloid. Such modes are essential to describe a quantized field in the Euclidean vacuum state in the open chart of de Sitter space. The contribution from these modes are noteworthy at small $l$ ($l \lesssim 10$) if the universe is well dominated by curvature at present. To compare our results with the previous analyses, we have calculated the same quantities for the conformal vacuum state, which have been adopted as a working hypothesis previously (Lyth & Stewart 1990; Rattra & Peebles 1994b; Kaminkowski et al. 1994). We have found that the difference appears for $l \lesssim 10$ when the spatial curvature is significant, e.g., $\Omega_0 \lesssim 0.1$. The possibility of the obsevational detection of this differences, however, will depend on the value of $\Omega_0$, and will be a delicate problem due to the cosmic variance and the obervational noise.

On the angular scales smaller than $l \sim 10$, the results will be same as the previous investigations. On those scales the contributions from the discrete modes are negligible as shown in Figures 2-6. As for the spectrum of the normalizable modes with $p^2 \geq 0$ the difference between the scalar field in the Euclidean vacuum and that in the conformal vacuum is the factor $\coth \pi p$ in Eq.(3.16), which is almost unity for $p \gtrsim 1$. Hence, irrespective of the choice of vacuum, the spectrum of the density perturbation becomes scale invariant for $p \gg 1$.

As mentioned before, in this paper we have neglected the effect of the field excitation due to the tunneling process during the bubble nucleation. Whether this assumption is valid or not will depend on the details of potential models of the bubble nucleation. Therefore the quantum fluctuations of a field during and after the bubble nucleation process should be investigated in various situations including the effect of gravity (Tanaka & Sasaki 1994). We have already investigated such effects for the bubble nucleation process which occurs in the Minkowski background (Yamamoto et al. 1994; Hamazaki et al. 1995). The results show that the field fluctuations generated on scales larger than that of the curvature can be large. Extrapolating this result to the case of the de Sitter backgroud, we expect that the effect of the field excitation due to the tunneling process will appear on scales larger than the curvature scales as well. To test the viability of the present scenario, more detailed investigations are necessary.

We are grateful to Professor H. Sato for his comment and encouragement. We also thank N. Gouda and N. Sugiyama for comments. This work was supported in part by Monbusho Grant-in-Aid for Scientific Research Nos. 2841, 2010 and 05640342 and the Sumitomo Foundation.